\documentclass[
reprint,
nofootinbib,
amsmath,amssymb,
aps,
prl,
twocolumn,
floatfix,
]{revtex4-2}

\usepackage{amssymb}   
\usepackage{amsfonts}
\usepackage{amsmath}
\usepackage{mathtools}
\usepackage[mathscr]{eucal}

\usepackage{graphicx}
\usepackage{xcolor}

\usepackage{tikz}
\usetikzlibrary{quantikz2}

\DeclareUnicodeCharacter{FF0C}{}

\usepackage{hyperref}
\definecolor{bblue}{RGB}{0, 126, 204}
\definecolor{dgreen}{RGB}{0,102,51}
\definecolor{blueus}{RGB}{10, 49, 97}
\definecolor{teal}{RGB}{0,128,128}
\definecolor{orangered}{RGB}{255,69,0}
\hypersetup{
	colorlinks=true,       
	linkcolor=purple, 
	citecolor=bblue,        
	filecolor=magenta,      
	urlcolor=bblue           
}

\newcommand{\op}[1]{\hat{#1}}
\newcommand{\opH}{\widehat{H}}
\newcommand{\eq}[1]{\begin{align}#1\end{align}}

\newcommand{\note}[1]{{\color{black} #1}}

\begin{document}

\title{Matrix product state ansatz for the variational quantum solution of the Heisenberg model on Kagome geometries}

\author{Younes Javanmard}
\email[]{younes.javanmard@itp.uni-hannover.de}
\author{ Ugne Liaubaite}
\email[]{ugne.liaubaite@itp.uni-hannover.de}
\address{Institut f\"ur Theoretische Physik, Leibniz Universit\"at Hannover, Appelstra{\ss}e 2, 30167 Hannover, Germany}

\author{Xusheng Xu}
\affiliation{Department of Physics, State Key Laboratory of Low-Dimensional Quantum Physics, Tsinghua University, Beijing 100084, China}

\author{ Man-Hong Yung}
\address{Shenzhen Institute for Quantum Science and Engineering, Southern University of Science and Technology，Shenzhen, 518055, China.}
\affiliation{International Quantum Academy, Shenzhen, 518048, China.}
\affiliation{Guangdong Provincial Key Laboratory of Quantum Science and Engineering, Southern University of Science and Technology, Shenzhen, 518055, China.}
\affiliation{Shenzhen Key Laboratory of Quantum Science and Engineering, Southern University of Science and Technology, Shenzhen,518055, China.}

\author{Tobias J. Osborne}
\email[]{tobias.osborne@itp.uni-hannover.de}
\address{Institut f\"ur Theoretische Physik, Leibniz Universit\"at Hannover, Appelstra{\ss}e 2, 30167 Hannover, Germany}

\date{\today}

\begin{abstract}
The Variational Quantum Eigensolver (VQE) algorithm, as applied to finding the ground state of a hamiltonian, is particularly well-suited for deployment on noisy intermediate-scale quantum (NISQ) devices. Here we utilize the VQE algorithm with a quantum circuit ansatz inspired by the Density Matrix Renormalization Group (DMRG) algorithm. To ameliorate the impact of realistic noise on the performance of method we employ zero-noise extrapolation. We find that, with realistic error rates, our DMRG-VQE hybrid algorithm delivers good results for strongly correlated systems. We illustrate our approach with the Heisenberg model on a Kagome lattice patch and demonstrate that DMRG-VQE hybrid methods can locate, and  faithfully represent the physics of, the ground state of such systems. \note{Moreover the parameterized ansatz circuit used in this work is low depth and requires a reasonably small number of parameters, and so is efficient for NISQ devices.}
\end{abstract}
\maketitle

\section{Introduction}

Complex quantum systems play a central role throughout the sciences, and have numerous  applications, from the study of molecular structure for quantum chemistry through to materials design. Despite their importance, however, progress on understanding their physics faces continuing challenges. This is due, in no small part, to the exponentially growing dimension of hilbert space (as a function of particle number). This renders direct study via exact diagonalization effectively useless, except for systems of only a few particles. Instead one must take employ either perturbative, Monte Carlo sampling, or variational methods to make progress.

With the availability of near-term quantum information processing devices, new possibilities for the simulation of complex quantum systems have emerged. A key new technique available with quantum computation is the ability to directly simulate the Schrödinger equation itself. Here there has been dramatic progress with a multitude of available methods, including, Lie-Trotter expansions \cite{zalkaSimulatingQuantumSystems1998,childsTheoryTrotterError2021,lowComplexityImplementingTrotter2022} and more advanced techniques such as the quantum singular value transform \cite{gilyenQuantumSingularValue2019,martynGrandUnificationQuantum2021}, quantum walk methods (qubitization) \cite{lowHamiltonianSimulationQubitization2019,lowOptimalHamiltonianSimulation2017}, linear combination of unitaries \cite{childsHamiltonianSimulationUsing}, and randomized evolutions (e.g., qDRIFT \cite{campbellRandomCompilerFast2019} and density matrix exponentiation \cite{lloydQuantumPrincipalComponent2014a}). The success of these approaches are, however, limited to quantum computers with access to ideal error-free \emph{logical qubits}.

Presently available quantum information processing devices operate below the fault-tolerance threshold, so that the aforementioned methods cannot yet be employed. Instead, the current generation of noisy intermediate scale quantum (NISQ) devices \cite{preskillQuantumComputingNISQ2018} must take resort to approximate and variational methods. There has been, in the past years, intense activity in the development of approximate variational quantum methods. The most prominent NISQ-compatible approach is epitomised by the quantum approximate optimization algorithm (QAOA) \cite{farhiQuantumApproximateOptimization2014}. This ansatz, directly inspired by the adiabatic quantum algorithm \cite{farhiQuantumAdiabaticEvolution2001}, has received considerable interest since its appearance in 2014, and has been generalized in many directions, most notably via the quantum alternating operator ansatz \cite{hadfieldQuantumApproximateOptimization2019}. The QAOA is an example of a class of variational heuristics which exploit parametrized quantum circuits to provide feasible solutions to optimization problems exploiting a hybrid variational methodology. A key example of a variational quantum heuristic of direct relevance to this work is given by the variational quantum eigensolver (VQE) \cite{peruzzoVariationalEigenvalueSolver2014}.

The success of variational quantum methods such as the VQE face several significant challenges. These include, the limited size of presently available quantum information processing devices, the deleterious effects of decoherence, and the fact that classical optimization methods are significantly more advanced. The first two are constantly being investigated, in particular, various \emph{quantum error mitigation} techniques have been proposed to improve the performance of currently available hardware without the requirement of mid-circuit measurement and ancillary correction qubits. Some of the most well-known techniques include zero-noise extrapolation and probabilistic error cancellation \cite{Temme2017,LiBenjamin2017}. Progress on the third factor, however, has been comparatively slower because classical variational simulation is extremely mature: Here tensor networks usually offer the best results available, with accuracies routinely achieving machine precision \cite{bridgemanHandwavingInterpretiveDance2017}. Even in higher dimensions tensor networks offer a very compelling general purpose approach. 

Variational quantum algorithms cannot, at the present time, compete with the accuracy of tensor network methods. One crucial factor here is that variational quantum approaches naively attempt to carry out a minimization over an arbitrary parametrized quantum circuits (PQC), ignoring known structure about the problem. This situation is reminiscent of the development of tensor networks before the year 2000: Many arbitrarily chosen tensor-network architectures were proposed and failed, usually due to instabilities and the proliferation of local minima. During the last 30 years there has been excellent progress in understanding the physical requirements of a tensor-network simulation and this directly led to the development of highly optimized methods \cite{verstraeteDensityMatrixRenormalization2023}. Interestingly, there have been very few studies which leverage tensor-network technology in the context of variational quantum methods. At the very least, to improve on TN methods, one should first classically pre-optimize a tensor network, and then subsequently preload the TNS into a quantum computer before further optimization. In this way variational quantum methods would never underperform relative to the best classical methods. 

There is now a small literature exploring the VQE and the TN ansatz. One early work was \cite{dborinMatrixProductState2022}: here a circuit pre-training method based on matrix-product state machine learning methods was introduced, and it was demonstrated that it accelerates the training of PQCs for both supervised learning, energy minimization, and combinatorial optimization. Similarly, the paper \cite{huangTensorNetworkAssisted2022} introduced TN-assisted parameterized quantum circuits, concatenating a classical tensor network operator to a quantum circuit and hence effectively increasing the circuit expressivity without physically realizing a deeper circuit. Since the appended TN unitary is realized as a classical rotation of the Hamiltonian, the resulting TN-PQC can effectively increase the circuit depth and, thus, the expressivity without physically realizing a deeper circuit. Both of these studies were very promising, and it is clear that much further work is required to consolidate this idea. Since these initial investigations there have been several papers applying TN pre-training, mostly in the context of quantum machine learning \cite{khanPreoptimizingVariationalQuantum2023,shinAnalyzingQuantumMachine2023,rieserTensorNetworksQuantum2023,fanQuantumCircuitMatrix2023}. 

In this paper we consider hybrid classical-quantum variational methods which exploit an initial tensor network which is loaded into a quantum device as a PQC and then subsequently subjected to variational optimization. In order to improve the performance on NISQ devices error mitigation schemes are applied to extract observable properties. The scheme described here is applied to the Heisenberg model on a Kagome lattice patch, a challenge problem recently proposed by \cite{IBMQuantumAwards2023}. \note{Few studies used VQE to solve this problem by considering different ansatz and optimization methods\cite{Kattemolle2021, s2023efficient}. The investigation of magnetic materials and model systems showcasing quantum spin liquid behavior is currently a subject of considerable experimental and theoretical interest. Among these systems, the antiferromagnetic Heisenberg model on the Kagome lattice geometry stands out as a prototypical highly frustrated quantum magnet in two spatial dimensions, with the potential to exhibit spin liquid behavior. Nonetheless, attaining a thorough comprehension of this ostensibly straightforward model has proven unexpectedly challenging\cite{Balents2010}.}
\begin{figure*}[t!]
\centering
\includegraphics[width=\textwidth]{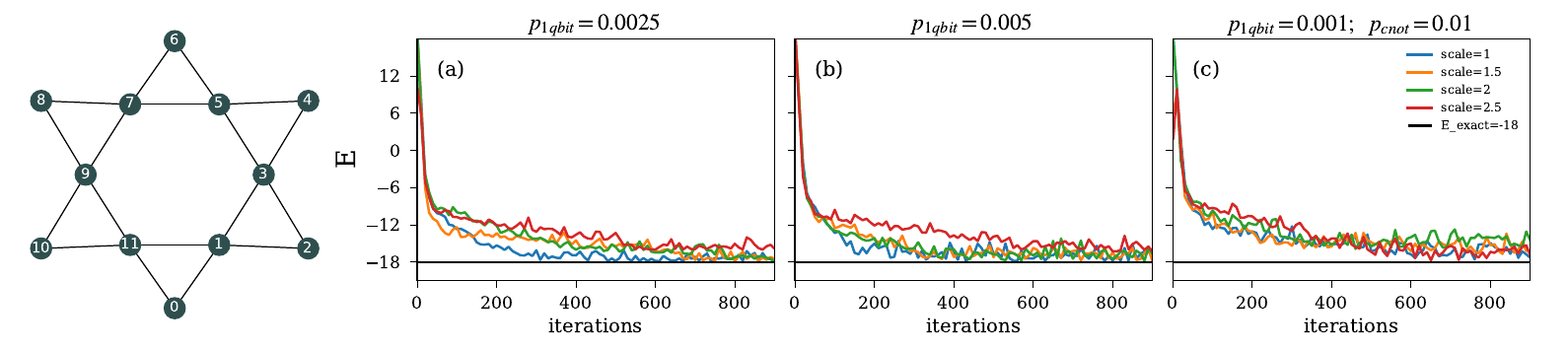}
\caption{(Left) Kagome star geometry with a zig-zag labelling defining our model. (Right) Hybrid DMRG-VQE results: The behaviour of energy as a function of number of iterations for VQE. (a). for one qubit depolarizing noise $p=0.025$. (b). For one qubit depolarizing noise $p=0.005$. (c). adding additional depolarizing noise to each cnot gate $p_{cnot}=0.01$ and for each one qubit gate $p=0.001$.}
\label{fig: Kagome star and results}
\end{figure*}

\begin{figure}[th!]
\centering
\includegraphics[width=\columnwidth]{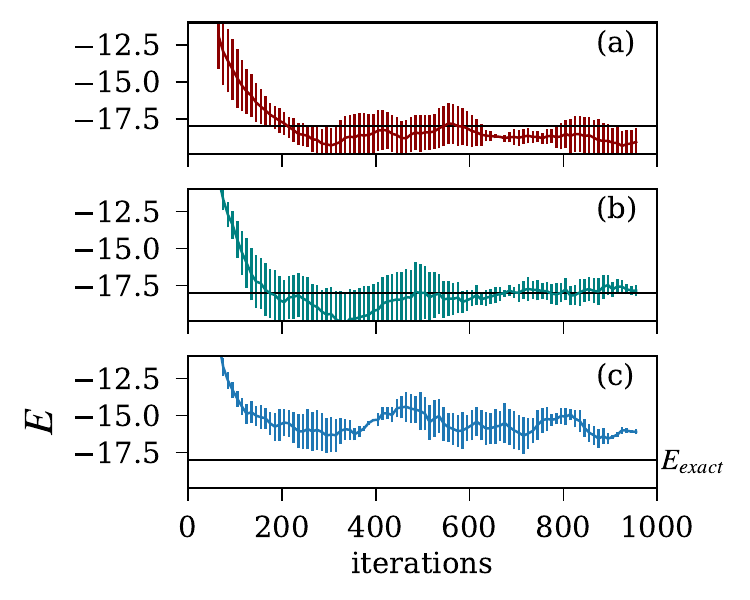}
\caption{\note{ZNE results for all iterations including the error-bars. Plots (a), (b) and (c) correspond to Fig.~\ref{fig: Kagome star and results}.}}
\label{fig: ZNE all iterations}
\end{figure}

\begin{figure}[th!]
\centering
\includegraphics[width=\columnwidth]{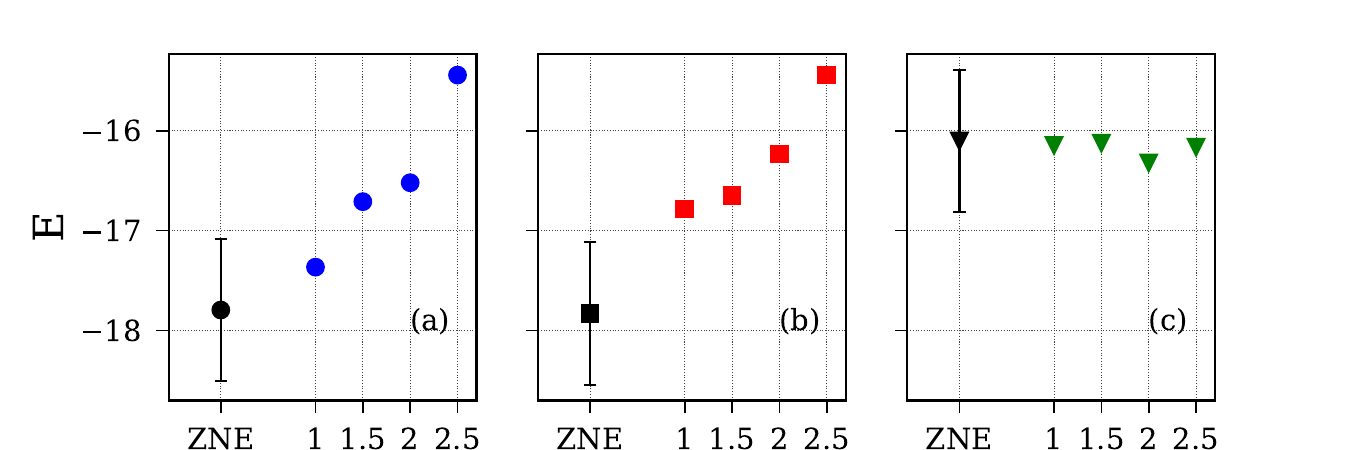}
\caption{ZNE results. Plots (a), (b) and (c) correspond to Fig.~\ref{fig: Kagome star and results}. \note{The data points corresponds to the last steps in the Fig.~\ref{fig: ZNE all iterations}}}
\label{fig: ZNE}
\end{figure}
\section{Preliminaries}
The simulation target we consider in this paper is a strongly interacting quantum spin system involving a collection of $N$ quantum spin-$\frac12$ degrees of freedom. Thus the kinematical degrees of freedom are captured by the Hilbert space furnished by $\mathcal{H}\cong \left(\mathbb{C}^{2}\right)^{\otimes N}$, which is conveniently identified with that for $N$ qubits. We choose as computational basis for a single qubit the eigenstates of the Pauli operator $\op{Z}$, denoted $|0\rangle$ and $|1\rangle$. We employ the  notation $\ket{\sigma_1 \cdots \sigma_i \cdots \sigma_N}$, $\sigma_j\in \{0,1\}$, for a general computational basis state of $N$ qubits.

The dynamics for the simulation target are generated by a locally interacting Hamiltonian $\opH$ involving nearest-neighbour interactions $h_{ij}$ between pairs of spins:
\begin{equation}
    \opH = \sum_{\langle i, j \rangle} h_{ij}.
\end{equation}
Here the notation $\langle i,j \rangle$ means that the sum is taken over only those spins $i$ and $j$ which are adjacent with respect to a given lattice structure. The operator $h_{ij}$ is given by a $4\times 4$ hermitian matrix acting nontrivially on only spins $i$ and $j$, i.e., tensored with the identity operator on the remaining spins. In the sequel we assume that the lattice structure is low dimensional. We model higher-dimensional lattice structures by identifying the spins with a one-dimensional chain with potentially long-range interactions. 

A hamiltonian of particular significance in the study of strongly correlated systems is the XXZ Hamiltonian, in the form of the Heisenberg model, with interaction term 
\eq{h_{ij} \equiv J(\op{X}_i \op{X}_j + \op{Y}_i \op{Y}_j + \op{Z}_i \op{Z}_j),}
where $J$ is the coupling strength (here $J=1.0$) and $\op{X}_i$, $\op{Y}_i$, are $\op{Z}_i$ are the Pauli matrices acting nontrivially on spin $i$. 

The method we describe below is generally applicable to any strongly interacting quantum spin system $\opH$. However, in order to exemplify our results, we take the particular case of the Heisenberg model on a Kagome lattice patch involving $12$ spins. This system, and the spin labelling employed for our investigations, is illustrated in the left panel of Fig.~\ref{fig: Kagome star and results}. One crucial observation at this stage is that the the efficacy of variational methods is strongly dependent on the labelling of the spins: The physical explanation is that nearest-neighbour entanglement in one labelling becomes nonlocal, and thus harder to represent with a one-dimensional ansatz, with respect to a different labelling. The $12$-qubit Heisenberg model on a Kagome lattice star was recently identified as a key challenge problem for VQE methods due to the frustrated and highly entangled nature of the ground state \cite{IBMQuantumAwards2023}. 

The goal of this paper is to obtain a good representation $\rho$, stored in a quantum register of $N$ qubits, of the ground state $|\Omega\rangle$ of $\opH$. This quantum state can then be subjected to measurements to obtain estimates for physical observables. Because of the effects of decoherence, our representation $\rho$ will necessarily be a mixed state, represented with a density operator. 

\section{Methods}
In this section we summarise the methods we employ to extract estimates for physical observables with respect to the ground state $|\Omega\rangle$ of the spin system $\opH$ on a NISQ device 

\subsection{Matrix product state algorithms}
A key input for our approach is White's \emph{Density Matrix Renormalization Group} (DMRG) \cite{schollwockDensitymatrixRenormalizationGroup2005,schollwockDensitymatrixRenormalizationGroup2011,whiteDensityMatrixFormulation1992}. This is a classical numerical method to obtain a variational representation of the ground state $|\Omega\rangle$ in the form of what is known as a \emph{Matrix Product State} (MPS), which is a state of the form \cite{fannesFinitelyCorrelatedPure1994}:
\eq{\ket{\psi} = \sum_{\substack{\sigma_1\hdots\sigma_N\\ \alpha_1,\ldots, \alpha_{N-1}}} {T^{\sigma_1}_{\alpha_0, \alpha_1} \hdots T^{\sigma_i}_{\alpha_{i-1}, \alpha_i}\hdots T^{\sigma_N}_{\alpha_{N-1}, \alpha_N}}\ket{\sigma_1 \cdots \sigma_i \cdots \sigma_N},}
where each $T^{\sigma_i}_{\alpha_{i-1}, \alpha_i}$ is a rank-$3$ tensor. Note that the indices $\alpha_{i-1}$ and $\alpha_i$ range from $1$ to the \emph{bond dimension} $\chi_i$ on site $i$. The maximum value of the bond dimensions is denoted $\chi_{\text{max}}$. We impose open boundary conditions by requiring that $\chi_0=\chi_N = 1$.

The DMRG is a class of numerical algorithm which proceeds by sequentially optimizing the variational degrees of freedom of an MPS, namely the rank-3 tensors $T^{\sigma_i}_{\alpha_{i-1}, \alpha_i}$. Abstractly, the DMRG works by carrying out the quadratic variational optimization
\begin{equation}
    \min_{T^{\sigma_i}_{\alpha_{i-1}, \alpha_i}} \langle \psi |\opH|\psi \rangle,
\end{equation}
for each $i$, and iterating until convergence is reached. It is now known that MPS provide a faithful representation for the ground state of gapped strongly correlated system \cite{verstraeteMatrixProductStates2006,hastingsAreaLawOnedimensional2007}, ensuring the general applicability of the DMRG.

\subsection{Variational quantum algorithms}
The VQE is a hybrid variational scheme which exploits a parametrized quantum circuit to prepare trial wavefunctions which are then used to approximate the energy of the system. These trial states are optimized using classical optimization methods to approximate the ground state $|\Omega\rangle$ of a quantum many body system.

We denote the trial states for the VQE as $\ket{\bm{\theta}} \equiv U(\bm{\theta}) \ket{0}$, where the initial state $\ket{0}$ is a conveniently chosen product state (usually the ``all zeros'' state), to which a unitary PQC $U(\bm{\theta}) \equiv U_L(\theta_L) \cdots U_2(\theta_2)U_1(\theta_1)$ is applied. We assume that the PQC is built from a product of $L$ quantum gates $U_j(\theta_j)$, each of which depends on a parameter $\theta_j$. The energy expectation value $E(\bm{\theta}) \equiv \bra{\bm{\theta}} H \ket{\bm{\theta}}$ is then estimated via measurement on the quantum computer giving rise to an empirical approximation $\hat{E}$ to $E(\bm{\theta})$. This approximation is then supplied to a classical optimiser which updates the parameter values $\bm{\theta} \mapsto \bm{\theta} + \delta \bm{\theta}$ to minimise $\hat{E}$. This procedure is then iterated until sufficient convergence has been obtained, e.g.\ some threshold for the energy differences is surpassed. The ansatz state $\ket{\bm{\theta}}$ that achieved this threshold is then left in the quantum register and furnishes the variational approximation to the ground state $|\Omega\rangle$. Arbitrary observables for the variational representation $\ket{\bm{\theta}}$ can be decomposed into basic measurements of Pauli strings. Note that, on account of the probabilistic nature of quantum mechanics and the fundamental fluctuations encountered in estimating expectation values, the estimator $\hat{E}$ is a random variable, and the VQE is a probabilistic algorithm.

\subsection{A problem-efficient ansatz}
The effectiveness of the VQE algorithm crucially relies on the selection of an appropriate PQC ansatz for the trial wavefunctions. Poor choices of PQC will lead to spurious local minima (so-called ``barren plateaus'' being one manifestation \cite{McClean2018BarrenPlateausInQuantumNeuralNetworkTrainingLandscapes,Cerezo2021CostFunctionDependentBarrenPlateausInShallowParametrizedQuantumCircuits}) and inaccurate representation of the salient physics. This has of course been well understood in the tensor network literature for decades, and there has been tremendous progress in identifying appropriate variational states for strongly correlated systems. 

Here we leverage the learnings gained in the application of the DMRG to quantum spin systems by exploiting a general PQC construction to directly load an arbitrary MPS into a quantum register. This PQC was first explicitly described in 2007  \cite{schoenSequentialGenerationMatrixProduct2007}, and we summarise this construction here.  

\begin{figure*}[t!]
\centering
\includegraphics[width=\textwidth]{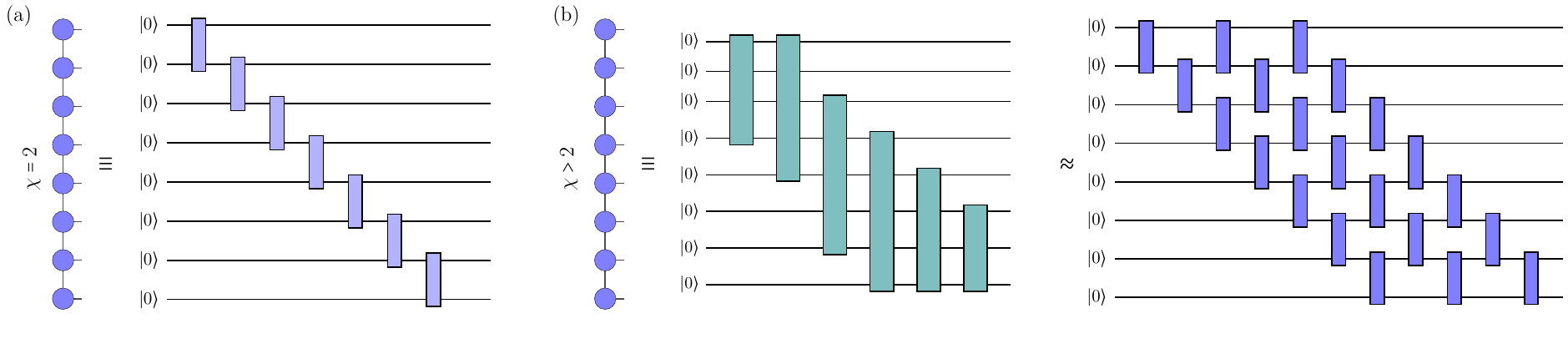}
\caption{Here we illustrate how to realise an arbitrary MPS via a unitary quantum circuit: \textbf{(a)} Exact mapping with bond dimension $\chi=2$; \textbf{(b)} When the bond dimension is larger than 2, the mapping is an approximation once the large unitaries are decomposed in terms of local $2$-qubit gates.}
\label{fig:mpspqc}
\end{figure*}

To describe the PQC we first ensure that the MPS is in \emph{right-canonical form}, by exploiting the gauge freedom of the MPS representation to enforce the additional conditions
\eq{\sum_{\sigma_i, \alpha_i} T^{\sigma_i}_{\alpha_{i-1},\alpha_i} T^{\sigma_i\dagger}_{\alpha_{i-1},\alpha_i} = I_{\alpha_{i-1}, \alpha_{i-1}}.} The right-canonical condition ensures that each individual tensor $B^{\sigma_i}$ is an isometry from $\ket{\alpha_{i-1}}$ to $\ket{\alpha_i, \sigma_i}$. Any isometry can be expressed as a unitary operation acting on an ancillary normalized state, denoted here as $\ket{0}_i$:
\eq{
T^{\sigma_i} &= U_i \ket{0}_i, \nonumber\\
T^{\sigma_i}_{\alpha_{i-1}, \alpha_i} &= \bra{\alpha_i, \sigma_i}U_i\ket{0_i, \alpha_{i-1}}.
\label{Eq: ismoetry}
}
In this way we can realise a general MPS $|\psi\rangle$ as a sequential staircase of $k$-local unitary operators $U_i$:
\eq{
    |\psi\rangle = U_N U_{N-1}\cdots U_1|0\rangle.
}
This construction is illustrated in Fig.~\ref{fig:mpspqc}. In the case the maximum bond dimension satisfies $\chi_{\text{max}}=2$ the unitaries are 2-qubit gates. We obtain our PQC by simply promoting all degrees of freedom of the unitaries $U_i$ to variational parameters. In the case of $\chi_{\text{max}}=2$ this results in at most $15\times (N-1)$ + $3$ parameters. If $\chi_{\text{max}}>2$ then the unitaries $U_i$ necessarily act on more than $2$ qubits at a time. In this case, for implementation reasons, we must compile the unitaries in terms of local $2$-qubit gates, further details are described in the Appendices. This introduces additional approximation errors.

Our procedure is then to apply the VQE directly to this PQC. If it were possible to calculate the required expectation values of the energy perfectly this would then lead to a variant of the DMRG. However, due decoherence and quantum projection noise, the estimators for the expectation values are inherently inaccurate. In order to cope with this we exploit error mitigation in the form of zero-noise extrapolation.

\subsection{Error mitigation}
Zero-noise extrapolation (ZNE), first introduced in \cite{LiBenjamin2017} and \cite{Temme2017}, makes use of the easy experimental accessibility of expectation values at different noise levels. This data is collected and subsequently extrapolated to provide an estimate for an ideal noiseless expectation value. The method is general, as it does not require any prior knowledge of the noise model.

To describe ZNE we explicitly denote the dependence of a noisy expectation value $\langle \op{O} \rangle_\lambda$ of an observable $\op{O}$ on the salient noise strength parameter $\lambda$. The ideal expectation value $\langle O \rangle _{\text{ideal}}$ is then the value of this function at $\lambda = 0$, i.e., $\langle O \rangle _{\text{ideal}} = \lim_{\lambda\rightarrow 0} \langle \op{O} \rangle_\lambda$. One can use a variety of extrapolation techniques to extract this limit from a sequence of expectation values. In this paper, we use \emph{Richardson extrapolation} to retrieve the noiseless expectation value, which is determined by 
\begin{equation}
    \langle O \rangle _{\text{ideal}} \approx \sum ^n _{k=1} \beta_k\langle O\rangle _{a_k\lambda}
\end{equation}
where $a$ are scaling factors arranged according to $1=a_0<a_1<a_2<...<a_n$. The coefficients $\beta$ are given as follows $\beta_k = \prod_{i\neq k} \frac{a_i}{a_k-a_i}$, which are obtained from the conditions that $\sum _{k=0}^n\beta_k = 1$ and $\sum^n_{l=0} \beta_l a^k_l = 0$.

Thus the crucial component of the ZNE procedure is the ability to scale the noise strength parameter $\lambda$. A common technique to achieve this, and the one we employ in this paper, is via so-called \emph{unitary folding}. The method is based on adding pairs of unitaries, consisting of the original unitary gate $U$ and its inverses $U^\dag$, to the circuit in the following manner: $U \longrightarrow U(U^\dagger U)^n$ where $n$ is an integer number. In the decoherence-free case the action of this new longer circuit is identical to the original. The scaling parameter $\alpha$ is the determined by $\alpha = 1+ \frac{2n}{d} $ where $d$ is depth of the circuit, and $n$ is natural number \cite{Giurgica_Tiron2020}. 

The dominant error after the ZNE procedure has been applied arises from higher-order terms neglected in the expansion of the expectation value in terms of the noise strength parameter $\lambda$. The cost of the mitigation protocol can be estimated on-the-fly by considering the increase in the variance related to the mitigation process and estimating the overhead on the shots required to achieve the variance of the original circuit. The cost of error mitigation techniques are discussed in detail in the following \cite{Giurgica_Tiron2020} and \cite{Takagi_2022}.

The entire mitigation scheme used in the paper, including circuit folding and Richardson extrapolation, is implemented in the MindSpore Quantum framework and is available as open source code in \cite{mq_2021}.  

\section{Results and discussion}
In this section we report on the performance of ZNE applied to the DMRG-VQE alorithm in the case of the Heisenberg model on a Kagome geometry comprising 12 qubits. This model is small enough that we can obtain the exact ground-state energy, yet it is large enough that the model exibits nontrivial frustration and entanglement. The exact ground-state energy is given by $E=-18$.

\subsection{Implementation details}
\begin{figure}[th!]
\includegraphics[width=\columnwidth]{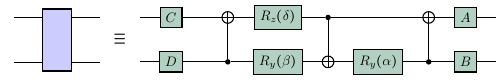}
\caption{Decomposing a general two-qubit gate in terms of at most three {\sc cnot} gates. Gates $A$, $B$, $C$ and $D$ are universal single-qubit rotation gate. In our work we consider them as $u3$-gates parametrized by three Euler angles: $u3(\theta, \phi, \lambda)$.}
\label{fig: two-qubit gate decomposition}
\end{figure}
The ansatz circuit we employ is equivalent to the one depicted in Figure~\ref{fig:mpspqc} (a). This circuit consists of a single layer of parameterized sequential two-qubit unitary gates, denoted as $U_i$. It's worth noting that any general two-qubit unitary gate can be decomposed into a maximum of three {\sc cnot} gates, as documented in the literature \cite{Shende2004, shende2004_2, Bremner2002}. \note{This decomposition is related to so-called A-gate  which preserve the symmetry of the system\cite{Gard2020}.} This decomposition is illustrated in Figure~\ref{fig: two-qubit gate decomposition}. In our parameterization, we introduce parameters for the single-qubit gates, as the {\sc cnot} gates are fixed. 

To faithfully replicate the quantum device's behavior in our simulations we exploit a noise model. This model accounts for the imperfections and errors that can occur during quantum computations. In our simulation, we model the noise using single-qubit and two-qubit depolarizing channels, which mimic the types of noise typically encountered in quantum hardware. The key parameter for these channels is denoted as `$p$' and represents the rate of depolarizing noise. The channels are thus described by the following equations:
\eq{
\mathcal{E}_{\text{1-qubit}}(\rho) &= (1-p)\rho + \dfrac{p}{3}(\op{X}\rho\op{X} + \op{Y}\rho\op{Y} + \op{Z}\rho\op{Z}) \\
\mathcal{E}_{\text{2-qubit}}(\rho) &= (1-p)\rho + \dfrac{p}{15}\sum_{\substack{\sigma^i, \sigma^j =[I,X,Y,Z]\\ \sigma^i=I, \sigma^j \neq I \\ \sigma^j=I, \sigma^i \neq I }} \sigma^i_1 \sigma^j_2 \rho \sigma^i_1 \sigma^j_2.\nonumber
\label{Eq: noise channels}
}
For the single-qubit noise channel, we consider three possible error cases, each having an equal probability of $p$. These error cases can be interpreted as the probabilistic application of Pauli gates $(X, Y, Z)$ to each qubit gate. In the case of the two-qubit noise channel, there are $15$ possible error combinations. Each combination shares the same error probability $p/15$. These combinations represent various error scenarios that can occur during the execution of a {\sc cnot} gate, taking into account the interaction of qubits and Pauli gates.

The introduction of noise in our simulations allows us to study the effects of these imperfections on the quantum algorithms and protocols we are analyzing, providing valuable insights into their robustness and performance in real-world quantum computing environments.
\subsection{Performance of the method}

We employ ZNE with linear extrapolation to acquire our results. To adjust the noise level, we randomly fold the circuit and utilize the scaling parameters $\alpha \in {1, 1.5, 2, 2.5}$. We have displayed the results in \note{Fig.~\ref{fig: ZNE all iterations} and \ref{fig: ZNE}} for each noise scaling, along with their corresponding ZNE values. When mitigating single-qubit depolarizing noise, our method achieves results very close to the exact solution. However, the precision of the results diminishes with the addition of two-qubit depolarizing noise, as the learnability of the ansatz parameters decreases.
\section{Nonlocal labeling}
\begin{figure}[th!]
\centering
\includegraphics[width=\columnwidth]{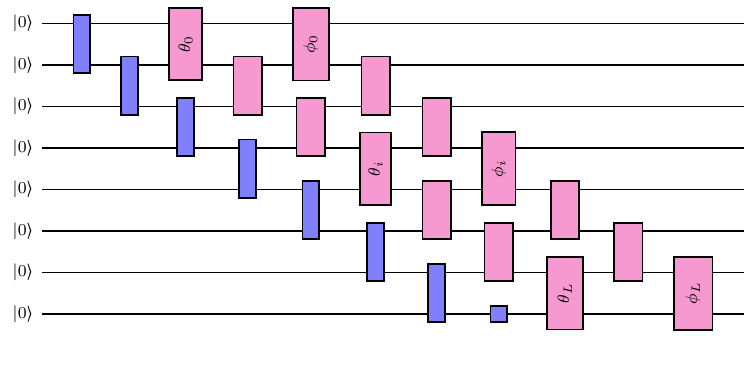}
\caption{MPS ansatz for non-local labeling of Kagome lattice. The quantum circuit used here has two parts. The first part is  a parameter-free optimized MPS with $\chi=2$ obtained by the DMRG and the second part is a parametrized quantum circuit based on the MPS tensor network. We consider each sequence layer corresponding to $\chi=2$ as one layer. This means the above quantum circuit has $D=3$ sequential layers. }
\label{fig: 3 sequential layers}
\end{figure}

To better illustrate the performance of the MPS ansatz used in this work, we change the labeling of the qubits in the Kagome lattice patch. This makes the problem artificially more challenging, as the entanglement structure of the ground state becomes nonlocal. This means that the simple MPS ansatz with $\chi=2$ is no longer able to faithfully represent the ground state. The ansatz must be improved to take account of the nonlocal entanglement. We enhance it by adding PQC layers, to take the form illustrated in Fig.~\ref{fig: 3 sequential layers}. To improve the performance of the VQE, we add a pre-trained parameter-free MPS circuit at the beginning of the ansatz (depicted in blue) and the rest of the sequential layers are parameterized and optimized via the VQE algorithm.

\begin{figure}[th!]
\centering
\includegraphics[width=1\columnwidth]{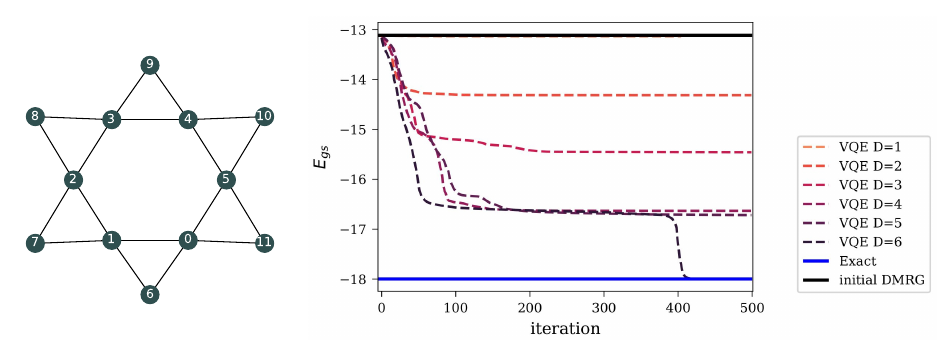}
\caption{Performance of the VQE for spiral (non-local) labelling for different sequential layers $D$. The optimization starts from pre-trained MPS with $\chi=2$ corresponding to Fig.\ref{fig: 3 sequential layers}.  }
\label{fig: veq result non-local}
\end{figure}

This type of inefficient labelling of the qubits would impact the performance of VQE. As one can see in the Fig.~\ref{fig: veq result non-local}, to achieve the same performance as for the zig-zag labeling in the main text, the inefficient labeling would need $D=6$ sequential layers of MPS ansatz.  

\section{Details on mapping a MPS tensor network to a parameterized quantum circuit}
\begin{figure}[th!]
\includegraphics[width=\columnwidth]{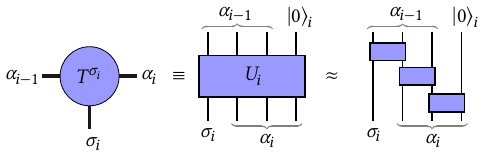}
\caption{Here we illustrate the steps to map a tensor for the MPS representation to its equivalent quantum gates which correspond to the isometric map in the main text.}
\label{fig: mps tensor to gates}
\end{figure}

Mapping an MPS tensor network to an equivalent quantum circuit has been recently discussed in applications of quantum computing \cite{Ran20, Lin21, Astrakhantsev23, Dborin_2022, rudolph2022decomposition}. Fig.~\ref{fig: mps tensor to gates} shows the how the isometry mapping for Eq.~\ref{Eq: ismoetry} works.
\section{Conclusions}
In this paper we have described a hybrid quantum algorithm based on the density matrix renormalization group and the variational quantum eigensolver. By realising a sequential unitary circuit to produce an arbitrary matrix product state we implemented a variational quantum method analogous to the DMRG directly to the MPS PQC. We exploited zero-noise extrapolation to overcome the effects of decoherence and quantum projection noise. Numerical experiments applied to the Heisenberg model on a Kagome geometry strongly indicate the expressivity and general utility of our method. 

There are a great variety of possible future directions to follow up on at this juncture. By exploiting the MPS ansatz and its many generalizations one should be able to extend the procedure described here to find variational quantum algorithms to approximate the dynamics of quantum spin systems, low-lying excited states, thermal states, and beyond, for both systems in one-dimension and higher.

\acknowledgements
Discussions with Soeren Wilkening are gratefully acknowledged.

\bibliography{references}
\bibliographystyle{apsrev4-1}

\appendix

\end{document}